# A New Lower Limit For The Bond Breaking Strains Of Defect-Free Carbon Nanotubes: Tight Binding MD Simulation Study

*Gülay DERELİ[*], Banu SÜNGÜ MISIRLIOĞLU, Önder EYECİOĞLU and Necati VARDAR*

*Department of Physics, Yildiz Technical University, Istanbul 34210, Turkey*

**Abstract**

The Order (N) Tight Binding Molecular Dynamics (TBMD) algorithms applied to simulate the tensile elongations of short (2-2.5 nm) armchair and zigzag Single Walled Carbon Nanotubes (SWCNTs) without bond breakings or defect formation. Simulations are repeated at high temperatures. We fix the lower limit of breaking strains to short SWCNTs without bond breaking or 5-7 defects formation. At room temperature, the simulated (4,4) SWCNT is able to carry the strain up to 130% of the relaxed tube length without bond breaking or 5-7 defects formation. This value is 127% for (11,0) SWCNT, 125% for (17,0) SWCNT, 123% for (10,10) SWCNT. Bond breakings occur at lower strain values in defect-free, short nanotubes as the radii of the nanotubes increase, regardless of their chirality. This is true even when we heat the tubes to higher temperatures. Bond breaking strain values, tensile strength, Young's moduli of the SWCNTs are obtained as functions of temperature. In general, defect-free zigzag nanotubes exhibit higher tensile strength than armchaired ones. Young's moduli of defect-free individual single-wall nanotubes are found to be in the range of 0.400 TPa within the elastic limit. At room temperature and experimentally realizable strain values, thinner tubes are more resistant to bond breaking and zigzag tubes over armchair ones. The same trend still holds at high temperatures although the resistance to strain gets lowered. We observe a slight decrease of the tensile strength with increasing temperatures. A similar behavior is also observed in Young's moduli. These results are important in determining the true breaking strains of SWCNTs.

[*]Correspondence should be addressed to gdereli@yildiz.edu.tr

# 1. Introduction

There has been increasing attention given to SWCNTs since they are the most durable material against stretching in cylinder's axial direction. The physical properties of SWCNTs are mostly determined by the chirality of the tubes. Depending on their chirality, the mechanical and electrical properties of SWCNTs change with strain and temperature [1-3]. Furthermore, their chirality distribution and their band gap can be tuned [4]. In recent years the outstanding mechanical, electrical and thermal properties of carbon nanotubes had been exploited to design nanodevices such as strain sensors [5,6] or to suggest new techniques for hydrogen production and storage [7]. Yet as another application, defect-free CNT fibers were considered for electrical wiring at nanoscales [8]. Currently, however, the applicability of CNTs is limited by a lack of control over their fundamental properties such as the diameter, length, and chirality. Especially for electronic applications, control over these properties is of crucial importance. It is well-known that defects also affect the electronic properties of SWCNTs in a drastic way. Electronic band structure of SWCNTs with defects show different behavior under elongation. Stretching a semiconducting SWCNT with defects causes narrowing of the energy band gap of the SWCNT. Thus, semiconductor to metal transitions are observed. Similarly for the metallic SWCNTs metal to semiconductor transitions are observed. This is a fundamental issue that remains to be solved during the elongations of SWCNTs with defects. Therefore study of the elongations of defect-free SWCNTs are especially important in electronic device applications of SWCNTs. Recent investigations on growing perfect SWCNTs to microscopic size have also attracted much attention. There are some new attempts in CNT fibre making [9] that exploit the unique yarn-like structure of defect-free CNT fibers to obtain exceptionally tough, resilient fibers against bending, with extremely high surface area [10]. Recent theoretical studies indicate that defect-free CNTs may be obtained with Fe/Ni/Co catalysts, see for instance Ref.[11].

In what follows, we model the tensile elongations of defect-free SWCNTs up to bond breaking. Our durable SWCNTs remain defect-free till the end of the elongation process allowing us to establish a lower limit for the bond breaking strains on defect-free SWCNTs.

There have been numerous computational and experimental studies aimed at the tensile elongation of SWCNTs. Critical failure strain of SWCNTs under tensile loadings were reported as 5-6 % or lower in the early experiments [12,13] and 30% in [14]. Their brittle and ductile behavior is reported in the high resolution TEM experiments of Margues et al. [15] and show that under high strain and low temperature conditions, (4,4) SWCNT, (4,0) SWCNT and (4,0)-(13,0) DWCNT are brittle. On the other hand, under low strain and high temperature, the tubes of diameter less than 1.1 nm show a completely ductile behavior, while larger tubes are moderately or completely brittle depending on their symmetry. Recent experiments of Huang et al. [16,17] reported the superplastic deformation behavior of CNTs. They can be elongated by more than 200% by simultaneously applying tensile stress and Joule heating. The interplay between the thermal energy and the mechanical strain energy speeds up the interactions among the defects. However, during such elongation processes SWCNTs shrink in diameter and due to the plastic flow of the defects mass loss might be observed. This has been revealed in Nanobeam Electron Diffraction (NBED) studies which indicate that the chiral indices of a nanotube decrease when tensile stress and electroresistive heating are simultaneously applied [18].

Tensile behavior of carbon nanotubes are subject to numerous computational studies. Several values of Young's modulus have been reported in the literature. The use of different procedures to represent the strain and also the use of different values for the wall thicknesses affect the results. For example, Hernandez et al. [19] gave values ranging from 0.78 to 1.26 TPa using tight binding simulations. On the otherhand, Sanchez-Portal et al. [20] obtained the Young's modulus values ranging from 0.50 to 0.80 TPa by using *ab initio* calculations.

Y. Wang et al. [21] show that the Young's modulus varies from 1.25 TPa to 1.48 TPa for the SWCNTs whose diameters range from 0.5 nm to 1.7 nm. The Young's modulus of a CNT decreases as the radius increases and the Young's modulus of a zigzag CNT is higher than that of an armchair CNT. This latter property is expected as the carbon atoms in a CNT are in $sp^2$ configurations and connected to one another by three strong σ bonds. Due to the geometric orientation of the carbon-carbon bonds relative to the nanotube axis, zigzag SWCNTs exhibit higher tensile strength and Young's moduli values compared to the armchair SWCNTs. In general, the Young's moduli depend more on the radii than on the helicities. These studies also show that the value of critical breaking strain increases as the radius decreases. These results are consistent with the results of A. Pullen et al. [22].

On the otherhand, C.-L. Zhang and H.-S. Shen [23] found that in the temperature range from 300K to 1200K, the values of Young's modulus decrease with increasing temperature and the increasing tube radius, for both armchair and zigzag nanotubes.

In our previous work, we have shown that the temperature variations have significant effects on stability and mechanical properties of (10,10) SWCNTs [24,25]. We observed that bond breaking strain values decrease with increasing temperature under elongation [24]. In the temperature range from 300 to 900K, both the Young's modulus and the tensile strength decrease with increasing temperature. At higher temperatures due to the deformation and softening of the SWCNT, Young's modulus starts to increase slightly while the tensile strength decreases.

The above results model elongations up to bond breakings. Results of similar studies in the literature differ depending on whether the tubes are defect-free or not during the process of elongation. Zhao et al. [26] calculated an activation barrier for 5-7 defect formation of 1.95 eV at 15% strain, making SWCNTs very stable against defect formation at room temperature. Recently, C.Tang et al. [27,28] simulated a superelongation mechanism of carbon nanotubes at high temperatures. Their results showed that tensile ductility exists in SWCNTs with large diameters (above that of the (31,0) tube) over 500-2400K temperature range. The interaction and evolution of defects that are already present generate new types of larger defects dominating the tensile elongation and lead eventually to the breaking of the SWCNT. This result provides important insights into the superelongation mechanism in nanostructures in general. However, the fundamental issue concerning the change in the electronic properties of the SWCNTs during elongation process remains unresolved. Defects change the electronic properties of SWCNTs. Electronic band structure of SWCNTs with defects show different behavior under elongation. Stretching a semiconducting SWCNT with defects causes narrowing of the energy band gap of the SWCNT. Thus, semiconductor to metal transitions are observed. For the metallic SWCNTs metal to semiconductor transitions are observed. This is a fundamental issue that remains to be solved during the elongations of SWCNTs with defects. Therefore study of the defect-free SWCNTs are important in electronic device applications of SWCNTs. To fully utilize metallic SWCNTs as interconnects in nanoscale electronic devices an ideal, structurally perfect single-walled carbon nanotubes should be kinetically stable and resist strains well beyond the 5-6 % observed experimentally. Recent experiments [29] using ultralong SWCNTs succeeded to reach strain values up to 13.7±0.3% without breakage or defect formation at room temperatures.

In the present work, we perform Tight Binding Molecular Dynamics (TBMD) simulations to model the tensile elongations of defect-free SWCNTs up to bond breaking. Thus our durable SWCNTs remain defect-free till the end of the elongation process. We have elongated four different SWCNTs as listed in Table:1. Elongations of the SWCNTs are repeated at three different temperature values (300K, 900K and 1800K). The elongations are stopped at the first sign of bond breaking. We find that for defect-free short nanotubes, as the radii of the nanotubes increase, the bond breakings occur at lower strain values regardless of the chirality. This fact remains true even when we heat the nanotubes to high temperatures. We obtain the bond breaking strain values, tensile strengths, and the Young's moduli of the SWCNTs as functions of temperature. Thus we have set a new lower limit to the breaking strains of defect-free SWCNTs. Older limit was 13.7±0.3% for ultralong SWCNTs using the Raman spectra observations without the indication of chirality [29]. We achieved strains up to 30% in (4,4) SWCNT, 23% for (10,10) SWCNT, 27% for (11,0) SWCNT and 25% for (17,0) SWCNTs at room temperature without a change in the tube chirality.

## 2. Method

The energy versus strain curves in this work are obtained through the use of an Order(N), Tight Binding Molecular Dynamics (TBMD) simulation method we developed for SWCNTs in our earlier works [30-32]. TBMD is a computational tool designed to run finite temperature MD simulations within the semi-empirical tight binding scheme [33]. This technique can be used to simulate material systems at different conditions of temperature, pressure etc., including materials at extreme thermodynamical conditions. Simulations are performed in two steps: i) SWCNTs are simulated until their structural stability is sustained, ii) strain is applied and SWCNTs are simulated for another 2000 MD steps. At each MD step, total energy and atomic forces are calculated.

In TBMD, electronic structure of the system can be calculated from the TBMD Hamiltonian below so that the quantum mechanical many-body nature is taken into account [33]:

$$H_{TBMD} = \sum_{\alpha} \frac{p_{\alpha}^2}{2m_{\alpha}} + \sum_{n} \varepsilon_n f(\varepsilon_n, T) + U_{rep} \quad . \tag{1}$$

The forces needed to move atoms is evaluated from the TBMD Hamiltonian above as

$$\vec{f}_{\alpha} = -\sum_{n}^{\#} \left\langle \psi_n \left| \frac{\partial H}{\partial \vec{r}_{\alpha}} \right| \psi_n \right\rangle f(\varepsilon_n, T) - \frac{\partial U_{rep}}{\partial \vec{r}_{\alpha}} \tag{2}$$

where the second term on the right hand side of the Eq. (2) is a repulsive force that will be given analytically as a function of the interatomic distance [34] and the first term is the Hellmann-Feynman contribution to the total force

$$\sum_{n} \left\langle \psi_n \left| \frac{\partial H}{\partial \vec{r}_{\alpha}} \right| \psi_n \right\rangle f(\varepsilon_n, T) = -2 \sum_{n} f(\varepsilon_n, T) \sum_{l\gamma} \sum_{l'\beta} C_{l'\beta}^n \frac{\partial H_{l'\beta, l\gamma}(r_{\beta\gamma})}{\partial \vec{r}_{\alpha}} C_{l\gamma}^n \quad . \tag{3}$$

Total energy of the system of ion cores and valance electrons can be written as

$$E_{tot} = 2 \sum_{n} \varepsilon_n f(\varepsilon_n, T) + U_{ii} - U_{ee} = E_{bs} + U_{rep} \tag{4}$$

where $f(\varepsilon_n, T)$ is the Fermi-Dirac distribution function. The sum over all the single particle energies is commonly called the band structure energy $E_{bs}$.

In order to calculate the band structure energy and Hellmann-Feynman forces, we need the full spectrum of eigenvalues $\varepsilon_n$ determined by solving the secular Eq. (5) below and the corresponding eigenvectors $C_{l'\beta}^n$

$$\sum_{l'\beta} \left( \left\langle \varphi_{l'\beta} | H | \varphi_{l\alpha} \right\rangle - \varepsilon_n \delta_{ll'} \delta_{\alpha\beta} \right) C_{l'\beta}^n = 0 \tag{5}$$

We diagonalize the TBMD Hamiltonian matrix at every time step of the simulation. Standard diagonalization of the TB matrix requires a computation time in cubic scaling with respect to the number of atoms and dominates the overall computational workload of the TBMD simulations. On the other hand, Order(N) methods solve for the band energy in real space and make the approximation that only local environment contributes to the bonding and hence the band energy of each atom. In this case, the computation time would be in linear scaling with the respect to the number of atoms. One of the O(N) methods that is widely used to carry out quantum calculations is the divide-and-conquer (DAC) approach. The key idea of this approach is to provide a description of a large system in terms of contributions from subsystems. The system splits into subsystems described by a set of basis set of neighboring atoms [31]. Each subsystem is solved separately and the total electron density and the energy of the system are obtained by summing the corresponding contributions from all subsystems. Furthermore we efficiently parallelized our O(N) TBMD algorithms. Application of the O(N) technique and the parallelization scheme can be followed in our previous studies [30-32].

In this paper, we report O(N) TBMD simulations on tensile elongations of various SWCNTs at temperatures 300K, 900K and 1800K. In our simulations, isotropic strain is applied along the axial (z) direction and the SWCNT is allowed to relax along the radial direction in order to preserve the volume of the system. We observe vibrations along the radial direction in the transverse plane. This has been noted in one of our earlier study [32]. Throughout this procedure the volume of the tube is kept constant. Thus the axial and radial deformations are strongly correlated. Zigzag and armchair SWCNTs have been chosen because of their structural simplicity and suitability as limiting cases in a broader study of chiral tubes. In the zigzag (n,0) tubes, the two sides of each hexagon are parallel to the tubular axis, whereas in the armchair (n,n) tubes, the two sides of each hexagon are perpendicular to the tube axis. We have chosen (10,10) and (17,0) nanotubes which are having exactly the same radius but different chirality. On the other hand, for observing the effects of varying radii, we have chosen a (4,4) nanotube due to its small radius. We selected a (11,0) nanotube which has a larger radius than (4,4) but smaller than those of (10,10) and (17,0). Dimensions of the SWCNTs in the study are given in Table:1. A periodic boundary condition is imposed in the uniaxial direction. All the simulations presented here are carried out in the canonical (NVT) ensemble. The Newtonian equations of motion are integrated using the velocity Verlet algorithm with a time step equal to 1 fs. To avoid an inaccurate integration, the velocities of the constituent atoms are occasionally rescaled to maintain the temperature of the system at the target value. In our simulations we used thermal equilibrium method, which is first reported in reference [25]. Thermal equilibrium (fast heating) simulations bring SWCNTs to high temperatures rapidly. The general trend in thermal processing is to reduce the process temperature and duration as much as possible in order to restrict the motion of atoms through atomic diffusion. Fast thermal processing restricts the diffusion, which is important when the control of impurities in the process is important. Fast thermal processes are also important for semiconductor device technology. In order to mimic a fast heating thermal process, we relaxed the optimized tube at the target temperature for a sufficiently long time. This eliminates the possibility of the system to be trapped in a metastable state. Then the uniaxial strain is applied along the tube axis. The axial strain is obtained from the formula $\varepsilon = (l - l_0)/l_0$ where $l_0$ is the equilibrium length in the axial direction for unstrained SWCNT and $l$ is the corresponding length in the strained SWCNT.

In our simulations, isotropic strain is applied along the axial (z) direction and the SWCNT is allowed to relax along the radial direction in order to preserve the volume of the system. We observe vibrations along the radial direction in the transverse plane. This has been noted in one of our earlier study [32]. Throughout this procedure the volume of the tube is kept constant. The strain applied SWCNT is equilibrated for another 2000 MD steps. This way we equilibrate and thermostate the system after the application of strain. Here we only examine tensile elongations which are generated by the applied positive strain. We obtain total energy per atom values for each strain value. For higher strain values, irreversible structural changes occur and the bond breakings start. This process was clarified in detail for (10,10) nanotubes in our previous work [24]. After the determination of the bond breaking strain, the stress-strain curves, the tensile strength and Young's modulus are obtained for the elongated SWCNTs. The stress is determined from the resulting force acting on the tube per cross-sectional area under tensile elongation. The cross-sectional area of the tube is defined by $S=2\pi R\delta R$ where $R$ and $\delta R$ are the radius and wall thickness of the tube, respectively. We have used 3.4 Å for the wall thickness of the nanotube.

The tensile strength and Young's modulus are two of the important parameters that specify the behavior of a material under strain. The tensile strength can be defined as the maximum stress which may be applied to the tube without perturbing its stability. On the other hand, Young's modulus, which shows the resistivity of a material to a change in its length, is determined from the slope of the stress-strain curves. SWCNTs will be locally subjected to abrupt temperature increases in electronic circuits and the temperature increase affects their structural stability and the mechanical properties.

Under appropriate strain and temperature values armchair SWCNTs show a ductile behavior. In our previous work [24,25] we have shown that (10,10) carbon nanotubes are brittle up to 900K and ductile afterwards (1800K). So in this study, several classes of behavior of SWCNTs under low/high strain and at low/high temperature values have been identified through MD simulations. Strain is applied until bond breakings occur that are identified via the sharp peaks that develop in the total energy graphs. Structurally stable tubes are included in the total energy versus strain graphs. At low/high temperatures bond breaking strain values, stress-strain curves, tensile strength values, Young's moduli of the armchair and zigzag SWCNTs are displayed.

## 3. Results and Discussion

Fig.1 summarizes the strain scenario applied to short (2-2.5 nm) armchair and zigzag SWCNTs. i)Tubes are optimized in 3ps of simulation time (We have tested longer simulation times and the energy fluctuations remained unchanged). ii) Next, axial strain is applied in small increments and the tube is equilibrated for another 2 ps of simulation time under this strain. iii) Elongation is stopped after the first sign of bond breaking between the carbon atoms. Therefore the enduring nanotubes are elongated but remain defect-free. Snapshots of the simulations are given just before the bond breakings. We obtain total enegy values (eV/Atom) from these graphs. Simulated defect-free (4,4) SWCNT is able to carry the strains up to 130% of the relaxed tube length in elongation at 300K. This value is 123% for (10,10) SWCNT, 127% for (11,0) SWCNT and 125% for (17,0) SWCNT. As the temperature is increased to 900K, defect-free (4,4) SWCNT is able to carry the strain only up to 127% of the relaxed tube length in elongation. This value decreases to 114% for (10,10) SWCNT, 122% for (11,0) SWCNT and 116% for (17,0) SWCNT. At 1800K the simulated defect-free (4,4) SWCNT is able to carry the strain up to 116% of the relaxed tube length in elongation. This value is 108% for (10,10) SWCNT, 113% for (11,0) SWCNT and 110% for (17,0) SWCNT. Results show that defect-free SWCNTs with smaller radii are able to carry the strain more efficiently than those with larger radii and for the same radius value a defect-free zigzag SWCNT is enduring stress better than an armchaired one. Experimentally, ultralong suspended SWCNTs were stretched to 13.7±0.3% without breakage or defect formation at room temperatures [29]. We have achieved strains up to 30% in (4,4) SWCNT, 23% in (10,10) SWCNT, 27% in (11,0) SWCNT and 25% in (17,0) SWCNT at room temperatures with short (2-2.5 nm) SWCNTs.

Fig.2 shows the bond breakings occuring in (4,4) SWCNT at 1800K with 17% stretching. Since the SWCNTs are not tied to anywhere we mostly observe the bond breakings at the ends. SWCNTs are still defect-free after the first signs of bond breakings which are identified via the sharp peaks that develop in the total energy graphs.

In Fig.3(a-c) we present the total energy values (eV/Atom) as functions of strain at three different temperatures, namely 300K, 900K and 1800K. Small increments of tensile elongations has been applied until the bond breakings between the carbon atoms of the nanotubes are observed. Total energy (eV/Atom) increases as we increase the temperature. Total energy values align with respect to tubes radius value until the elastic limit (which is around 0.10 strain). Small radius tubes have larger total energy values. Chirality has no effect as long as the tubes are at the same radius. This can be seen in the total energy values of the (10,10) and (17,0) SWCNTs. Radii of the (10,10) and (17,0) SWCNTs are approximately the same. Their total energy values coincide upto 0.10 strain which is the elastic limit for these tubes. Beyond the elastic limit, the strain energies of zigzag SWCNTs increase. From the Fig.3(a) we can conclude that smaller the radius more durable is the tube against stretching and zigzag SWCNTs are more durable than armchair ones. This behavior is also observed in Fig. 3(b) and 3(c). This shows that heating the nanotubes to higher temperatures (900K, 1800K) does not change the bond breaking behavior of the nanotubes much when they are short and defect-free. On the otherhand over all the heating process decreases the stretching limit of defect-free SWCNTs.

Fig.4(a-c) shows the variations of bond breaking strain values with respect to tube radius at 300K, 900K and 1800K, respectively. At 300K, bond breaking in (4,4) SWCNT occurs at 0.31 strain, in (11,0)

SWCNT at 0.28 strain, in (17,0) SWCNT at 0.26 strain and in (10,10) SWCNT at 0.24 strain. In defect-free short nanotubes, as the nanotube's radius increases the bond breaking occurs at lower strain values regardless of the chirality. This is true when we heat the tubes to higher temperatures. In Fig.4(d) we present the bond breaking strain values with respect to temperature. Bond breakings of (4,4), (11,0), (10,10) and (17,0) SWCNTs with stretching is given at 300K, 900K and 1800K. Bond breaking strain value for short, defect-free (4,4) SWCNT decreases from 0.31 to 0.17 with heating. Similar behavior is also observed in defect-free (11,0), (10,10) and (17,0) SWCNTs. Bond breaking strain value for (11,0) SWCNT decreases from 0.28 to 0.14. Bond breaking strain value for a (17,0) SWCNT decreases from 0.26 to 0.11. On the otherhand, the bond breaking strain value for a (10,10) SWCNT decreases from 0.24 to 0.09. As the temperature increases, disintegrations of atoms from their locations is possible at lower strain values. This is due to the increase in the thermal motion of the atoms. TBMD scheme derives the interatomic forces governing the time evolution directly from the electronic structure of the simulated system. This way, increase in electronic temperature is also taken into consideration.

In Fig.5(a-d) we give the stress-strain curves of defect-free (4,4), (11,0), (10,10) and (17,0) SWCNTs at three different temperature values of 300K, 900K and 1800K, respectively. In Fig. 5(a), the stress-strain curves of a (4,4) SWCNT almost coincide at 300K and 900K. However, at 1800K the stress values increase. On the otherhand, in Fig.5(b) another armchair SWCNT, namely (10,10) with a larger radius exhibits a different behavior with increasing temperature. For this defect-free SWCNT, the stress value increases when we increase the temperature from 300K to 900K. But as we continue to heat the SWCNT to 1800K, the stress value drops. Snapshots indicate that bond breakings occur in such a way to unzip the SWCNT. In Fig.5(c) and Fig.5(d) we examine the two zigzag SWCNTs, namely (11,0) and (17,0) at three different temperature values. For these SWCNTs the stress increases as we increase the temperature. A closer look at the corresponding graphs shows that the zigzag tube (17,0) with a larger radius respond to strain with a larger stress at the temperature range from 300K to 900K, compared to another zigzag tube (11,0) with a small radius.

In Fig.6(a-c) tensile strengths of the four defect-free SWCNTs are presented at three different temperature values. Tensile strength is by definition the maximum value of the stress applied to the SWCNT without perturbing its stability. From the figures we see that the zigzag nanotubes exhibit higher tensile strength than armchaired ones and narrow tubes over the large radius tubes. This can be explained by the existence of strong C-C bonds aligned with the axis of the zigzag nanotube. In Fig.6(d) we summarize the temperature dependence of the tensile strength. In general, heating a nanotubes expected to decrease its tensile strength as observed in Fig.6(d). However, (4,4) SWCNT has very small radius (2.76 Å at 300K) and the application of a 0.27 strain at 900K reduces the radius to 2.46 Å (acts like a nanowire) and the tensile strength drops only slightly as temperature increases from 300K to 900K. At 300K, the tensile strength takes the value 93.69 GPa for (4,4) SWCNT, 105.56 GPa for (11,0) SWCNT, 98.11 GPa for (17,0) SWCNT and 83.23 GPa for (10,10) SWCNT. As the temperature increases to 900K and 1800K, tensile strength values of a (4,4) SWCNT decreases to 88.42 GPa and 66.85 GPa, respectively. Similar behavior is observed for (11,0), (17,0) and (10,10) SWCNTs. Tensile strength values of a (11,0) SWCNT decrease to 90.44 GPa and 71.94 GPa, respectively. For (17,0) SWCNT, they decrease to 77.45 GPa and 59.15 GPa, respectively. Finally for (10,10) SWCNT, they decrease to 77.45 GPa and 59.15 GPa, respectively.

In Fig.7(a-c), the Young's moduli of SWCNTs are calculated at three different temperature values from the slopes of the stress-strain graphs. It turns out that the zigzag SWCNTs (11,0) and (17,0) have higher Young's moduli than the armchaired ones (4,4) and (10,10). The Young's moduli of the zigzag SWCNTs decrease with increasing temperature. At 300K, Young's modulus of a (11,0) SWCNT is calculated as 105.56 TPa. As the temperature increase to 900K and 1800K, it decrease to 90.44 TPa and 71.94 TPa, respectively. Furthermore Young's modulus of a (17,0) SWCNT is calculated as 98.11 TPa at 300K. It decreases to 77.45 TPa at 900K and to 59.15 TPa at 1800K, respectively. However, this is not the case for armchair SWCNTs (4,4) and (10,10). In Fig.7(d), it is seen that the Young's modulus of a (4,4) SWCNT decreases from 0.410 TPa to 0.365 TPa. There is 12% decrease in Young's modulus when we increase the temperature from 300K to 900K. As we further increase the temperature to 1800K, we observe a slight increase in Young's modulus to 0.378 TPa (3.6%). Similar behavior is also observed in the other armchaired SWCNT (10,10) for which the Young's modulus drops from 0.401 TPa (300K) to 0.352 TPa (900K). As we further increase the temperature to 1800K, we observe a slight increase in Young's modulus to 0.365 TPa (3.7%). Thus we can conclude that defect-free armchair SWCNTs may still demonstrate some ductile behavior at 1800K.

**4. Conclusion**

It has been shown experimentally that it is possible to stretch SWCNTs over 200% with the help of defect nucleation and its motion at high temperatures [16,17]. There are several ways in which crystalline and amorphous materials can be engineered to increase their yield strength. By altering the dislocation density, impurity levels and/or the grain size (in crystalline materials), the yield strength of materials can be fine-tuned. Such fine-tuning typically produce defects, for instance dislocation impurities, in the materials. In order to remove such defects, a larger stress should be applied, thus causing a higher yield stress in the materials. Extensive MD examinations of the atomistic processes in SWCNTs associated with the defect nucleation and its motion at high temperature is given in Ref.[28]. However, recent experimental studies showed that ultralong, suspended SWCNTs are able to carry a stretch up to 13.7±0.3% without breakage or defect formation at room temperatures [29]. In this paper, we simulated a similar elongation process in short SWCNTs. We determined the bond breakings, before 5-7 defect formation, using O(N) TBMD algorithms. It is well-known that the TBMD scheme derives the interatomic forces governing the time evolution directly from the electronic structure of the simulated system. This approach gives a correct picture for the bond breakings between carbon atoms. We have explored here the tensile behavior and mechanical properties of SWCNTs depending on tube radius and tube structure. We achieved strains up to 30% in (4, 4) SWCNT, 23% for (10,10) SWCNT, 27% for (11,0) SWCNT and 25% for (17,0) SWCNTs at room temperature without a change in the tube chirality. Older limit was 13.7±0.3% for ultralong SWCNTs using the Raman spectra observations. This is significant in the nanoscale electronic device applications of the SWCNTs. The results are analyzed at the room temperature and two higher temperature values of 900K and 1800K. There are two pairs of armchair and zigzag nanotubes that we considered. For short SWCNTs, the bond breaking strain values drop slightly as the radii of the tubes increase. Mechanical properties of SWCNTs have been found to be sensitive to their radii and chirality. Generally, the bond breaking strain, Young's modulus and tensile strength for both armchair and zigzag SWCNTs show slight decrease as the tube radius is increased. Zigzag nanotubes exhibit higher tensile strength than armchaired nanotubes when they are subjected to the same type of tensile elongations. This latter property is expected as the carbon atoms in a CNT are in $sp^2$ configurations and connected to one another by three strong σ bonds. Due to the geometric orientation of the carbon-carbon bonds relative to the nanotube axis, zigzag SWCNTs exhibit higher tensile strength and Young's moduli values compared to the armchair SWCNTs. Young's modulus has a slight dependence on the tube structure. The Young's modulus of zigzag nanotubes exhibit slightly higher values than armchair nanotubes, especially at 300K and 900K. As the temperature further increases (1800K) Young's modulus values get closer to each other. However, the bond breaking temperature values are insensitive to the tube structure and show a similar dependence on radii for both armchaired and zigzag nanotubes. This behavior repeats at all temperature values we tested. In addition, we observe that all the mechanical parameters (i.e. the bond breaking strain, Young's modulus and tensile strength) decrease with increasing temperature as should be expected due to the thermal motion of the atoms.

In summary, we note that carbon nanotubes with smaller radii show more tensile stability over a wide range of temperatures. That is to say, structures with smaller radii are more stable and may carry higher mechanical parameter values. This result is consistent with the results of Refs.[21,22]. On the otherhand, the tensile strength and the Young's modulus of zigzag nanotubes exhibit slightly higher values than those of the armchaired nanotubes. This can be explained again by the presence of strong longitudinal carbon bonds along the tube axis in zigzag nanotubes. It is remarkable that the TBMD results of our work bring information about the tensile elongation of SWCNTs without any mass loss at elevated temperatures. They provide useful insights for understanding the tensile elongations without plastic deformations over a wide temperature range. In fact SWCNTs may benefit by a powerful self-repairing capability that would allow extensive deformations at high temperatures. However, we used to think that this could be achieved only with a cost of mass loss and unpredictable electronic behavior of SWCNTs. In this paper we were able to show that SWCNTs would be able to endure enormous strains at higher temperatures, without bond breaking or the formation of 5-7 defects even when they are only a few nanometers long. Our O(N) TBMD simulation method does not extend to the multi-walled CNTs so we cannot comment on those cases.

**Acknowledgments**

The research reported here is supported through the Yildiz Technical University Research Fund Project No: 2009-01-01-KAP01. GD and NV also acknowledge the support from YTU Research Fund Project No: 2010-01-01- DOP01. Simulations are performed at the "Carbon Nanotubes Simulation Laboratory" at the Department of Physics, Yildiz Technical University, Istanbul, Turkey.

http://www.yildiz/edu/tr/~gdereli/lab_homepage/index.html)

**Figure Captions:**

**Fig. 1.** Variation of the total energy (eV/Atom) with the applied strain: First 3000MD steps give the energy of the pristine SWCNT (without strain), next 2000 MD steps under the applied strain. Strain is applied in small increments until the first sign of bond breaking occurs. The indicated strain values show the lower limits for which the bond breakings start in short armchair and zigzag SWCNTs. Snapshots of the SWCNTs are just before the bond breakings occur. Simulations are performed at 300K, 900K and at 1800K.

**Fig. 2.** Bond breakings are observed between the carbon atoms of (4,4) SWCNT for the strain of 0.17 at 1800K.

**Fig. 3.** Total Energy (eV/Atom) versus strain curves of armchair and zigzag SWCNTs: (a) at 300K, (b) at 900K, (c) at 1800K.

**Fig. 4.** Bond breaking strain values of armchair and zigzag SWCNTs: (a) at 300K, (b) at 900K, (c) at 1800K; (d) Bond breaking strain values of the SWCNTs as function of temperature.

**Fig. 5.** Stress-strain curves of armchair and zigzag SWCNTs at different temperatures: (a) (4,4) tube, (b) (10,10) tube, (c) (11,0) tube, (d) (17,0) tube.

**Fig. 6.** Tensile strength of armchair and zigzag SWCNTs: (a) at 300K, (b) at 900K, (c) at 1800K; (d) Tensile strength asfunction of temperature.

**Fig. 7.** Young's modulus of armchair and zigzag SWCNTs: (a) at 300K, (b) at 900K, (c) at 1800K; (d) Young's modulus as function of temperature.

**Table: 1**

|                  | (4,4) | (11,0) | (17,0) | (10,10) |
|------------------|-------|--------|--------|---------|
| Number of Layers | 20    | 20     | 20     | 20      |
| Number of Atoms  | 160   | 220    | 340    | 400     |
| Radius (nm)      | 0.28  | 0.43   | 0.67   | 0.68    |
| Length (nm)      | 2.34  | 1.92   | 1.97   | 2.34    |

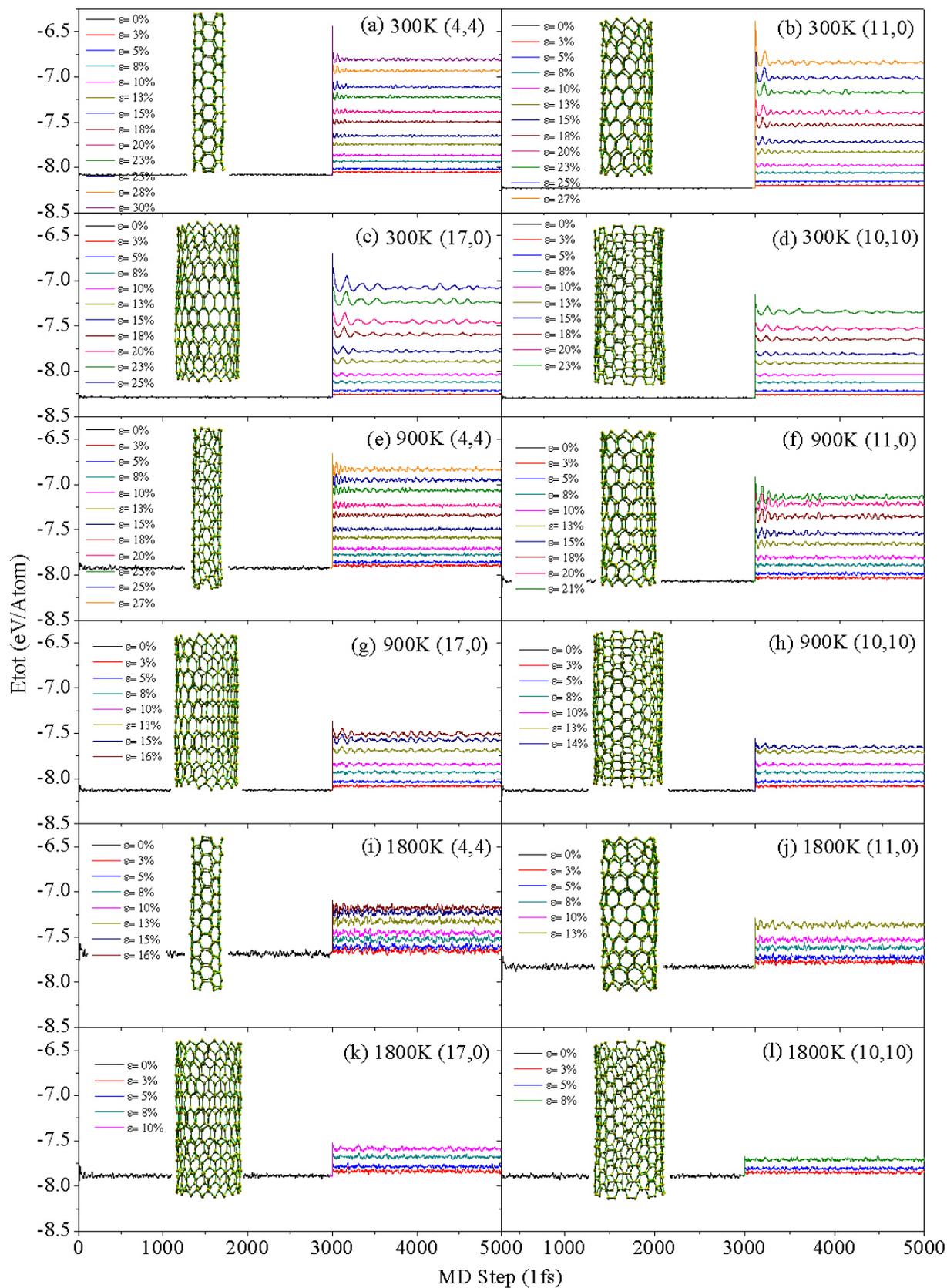

**Fig.1.**

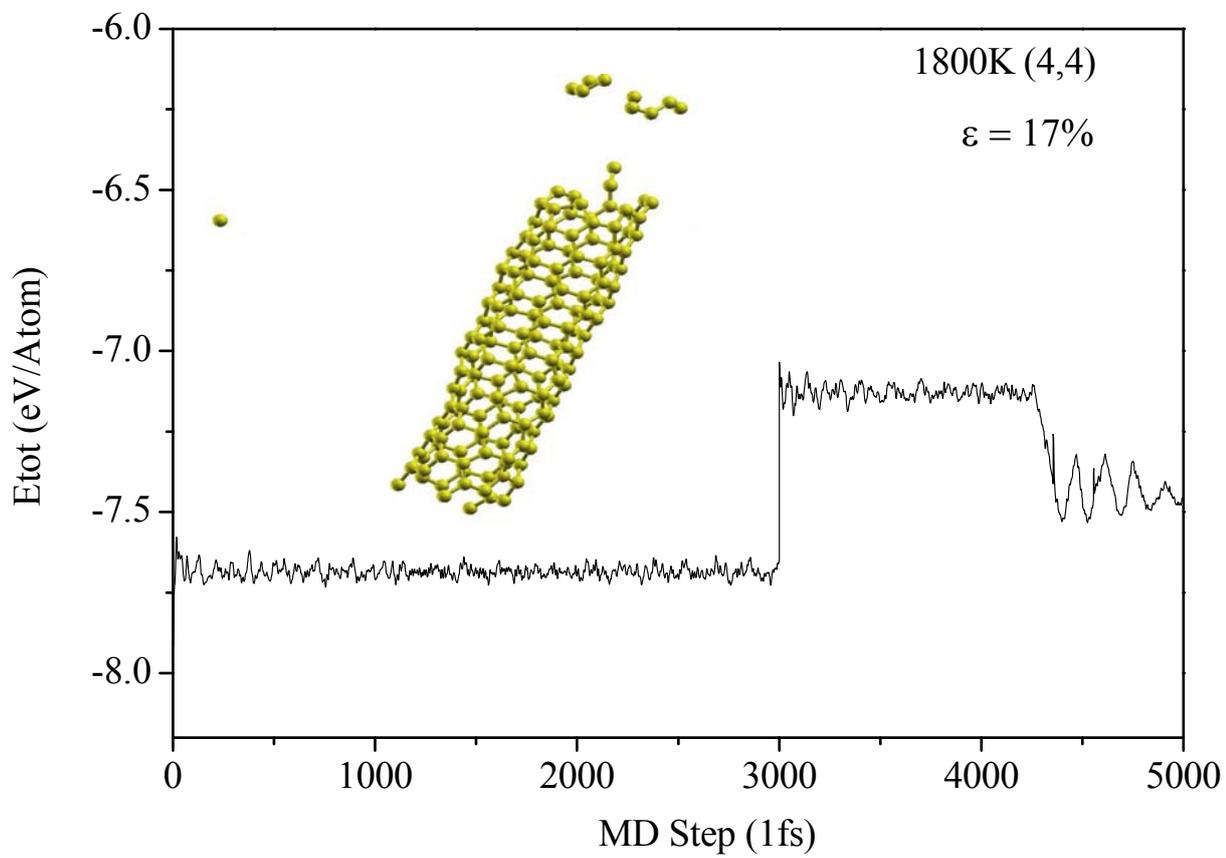

**Fig.2.**

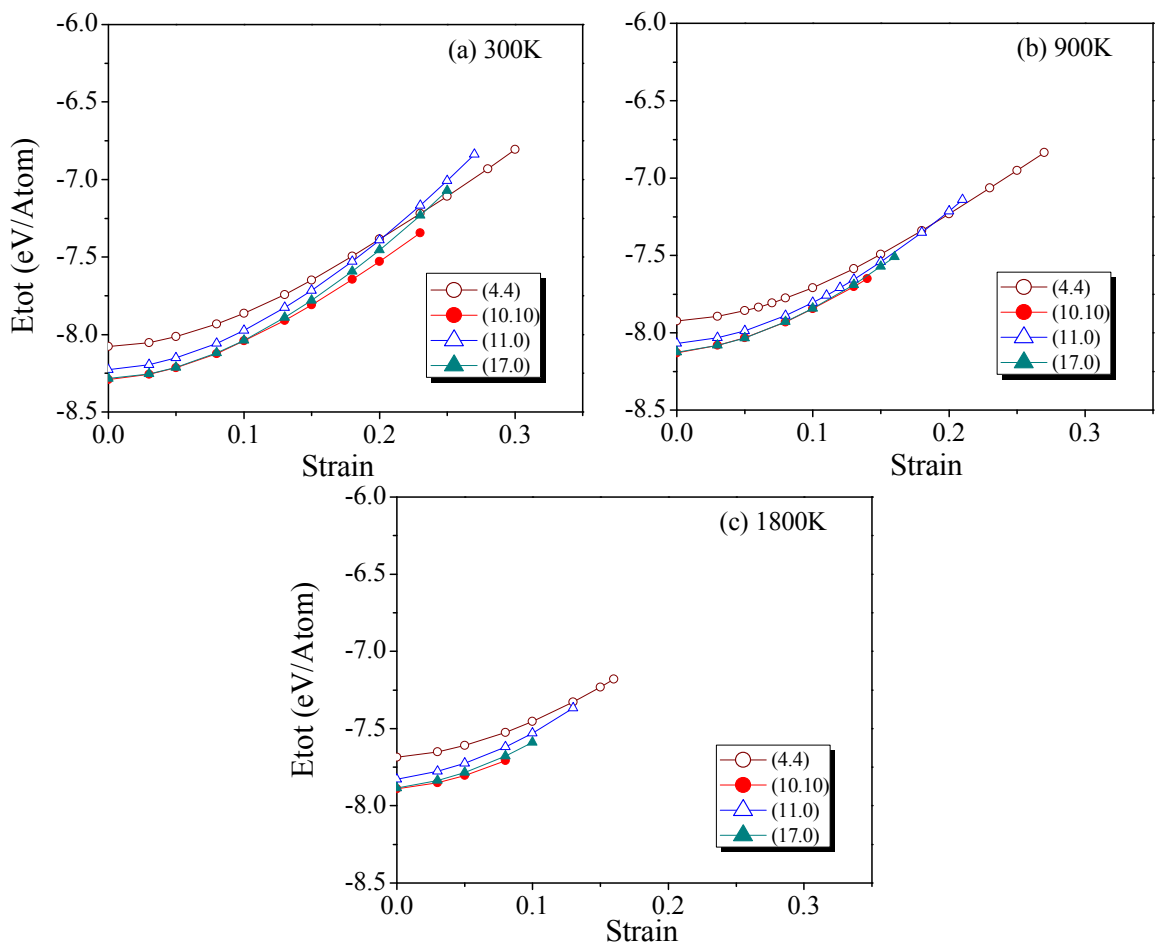

**Fig.3.**

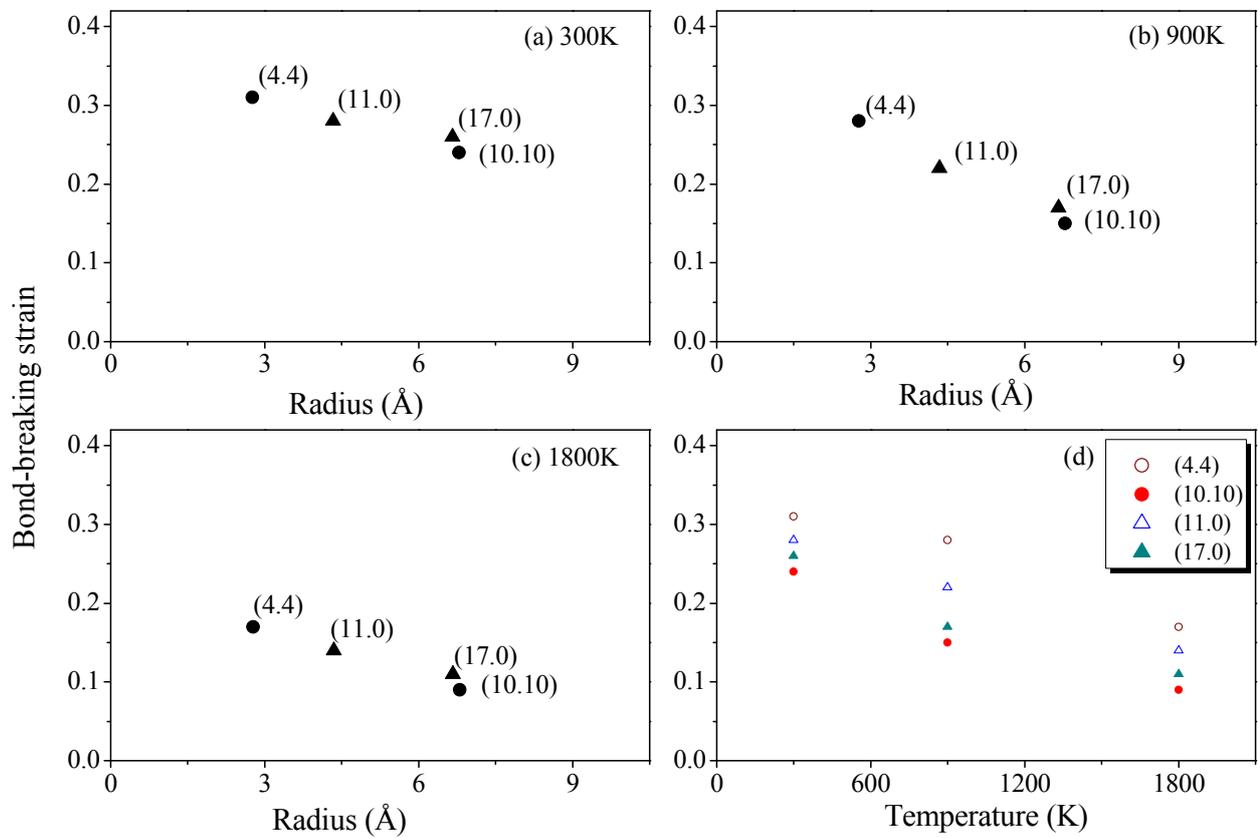

**Fig.4.**

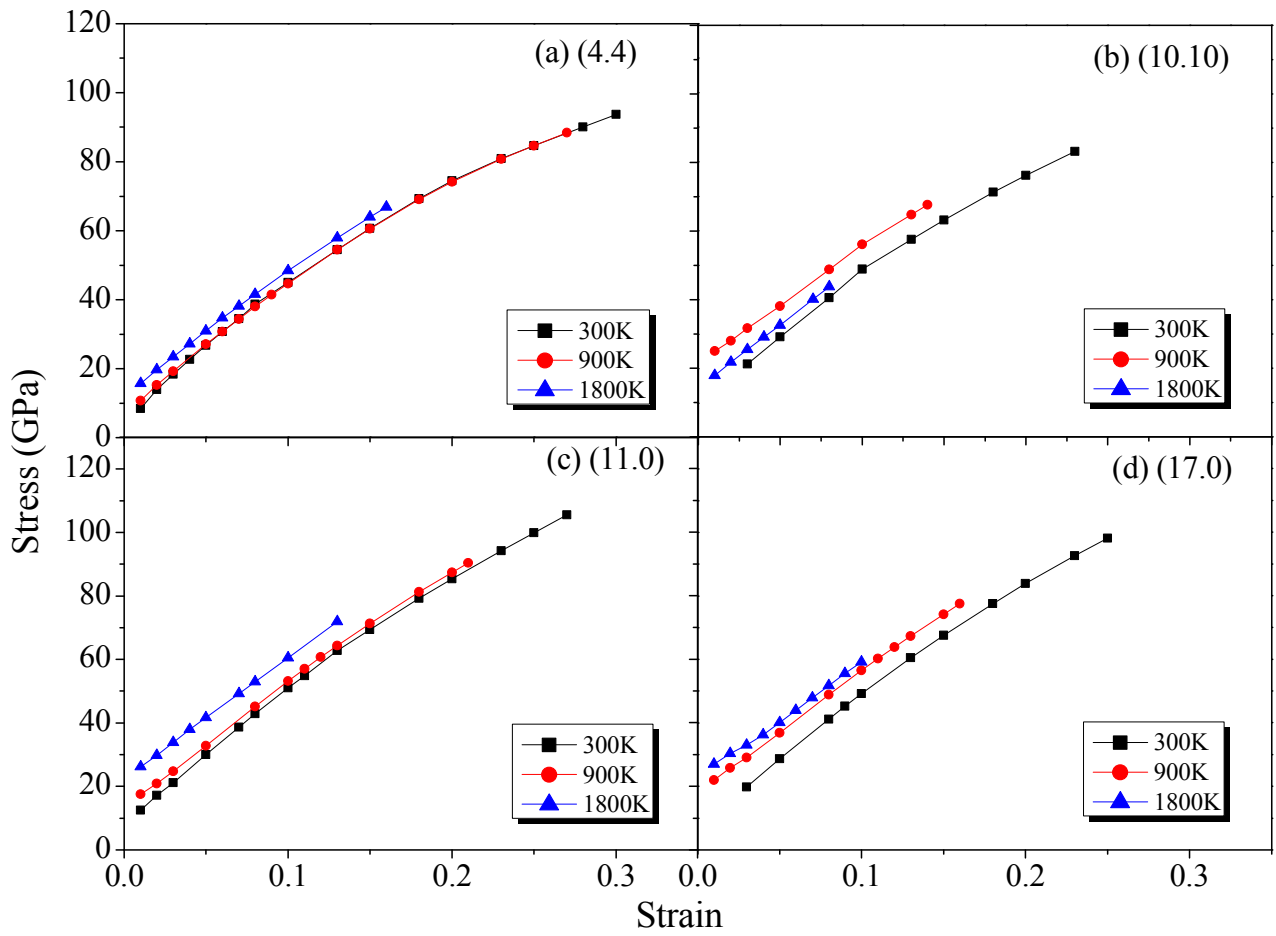

**Fig.5.**

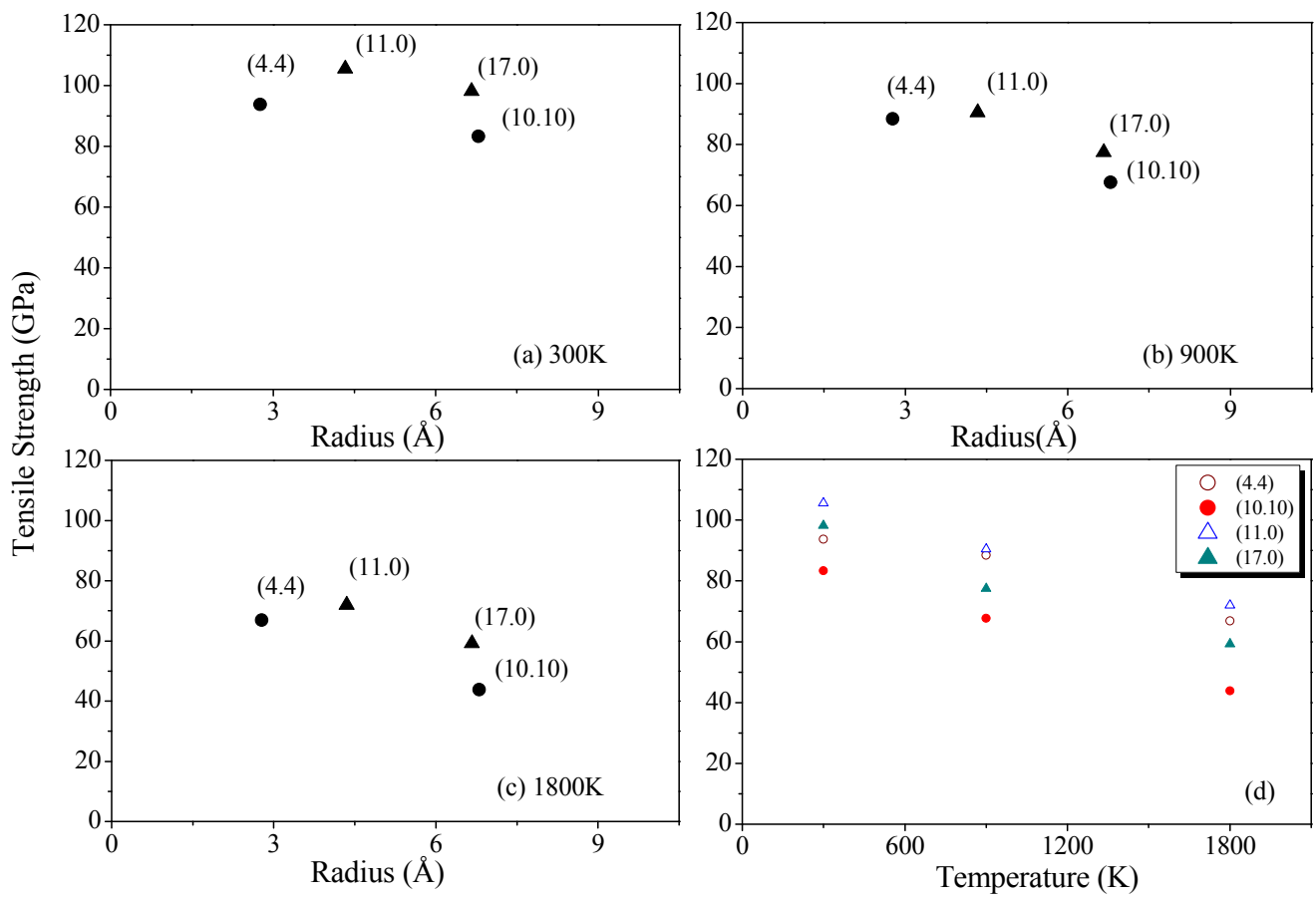

**Fig. 6.**

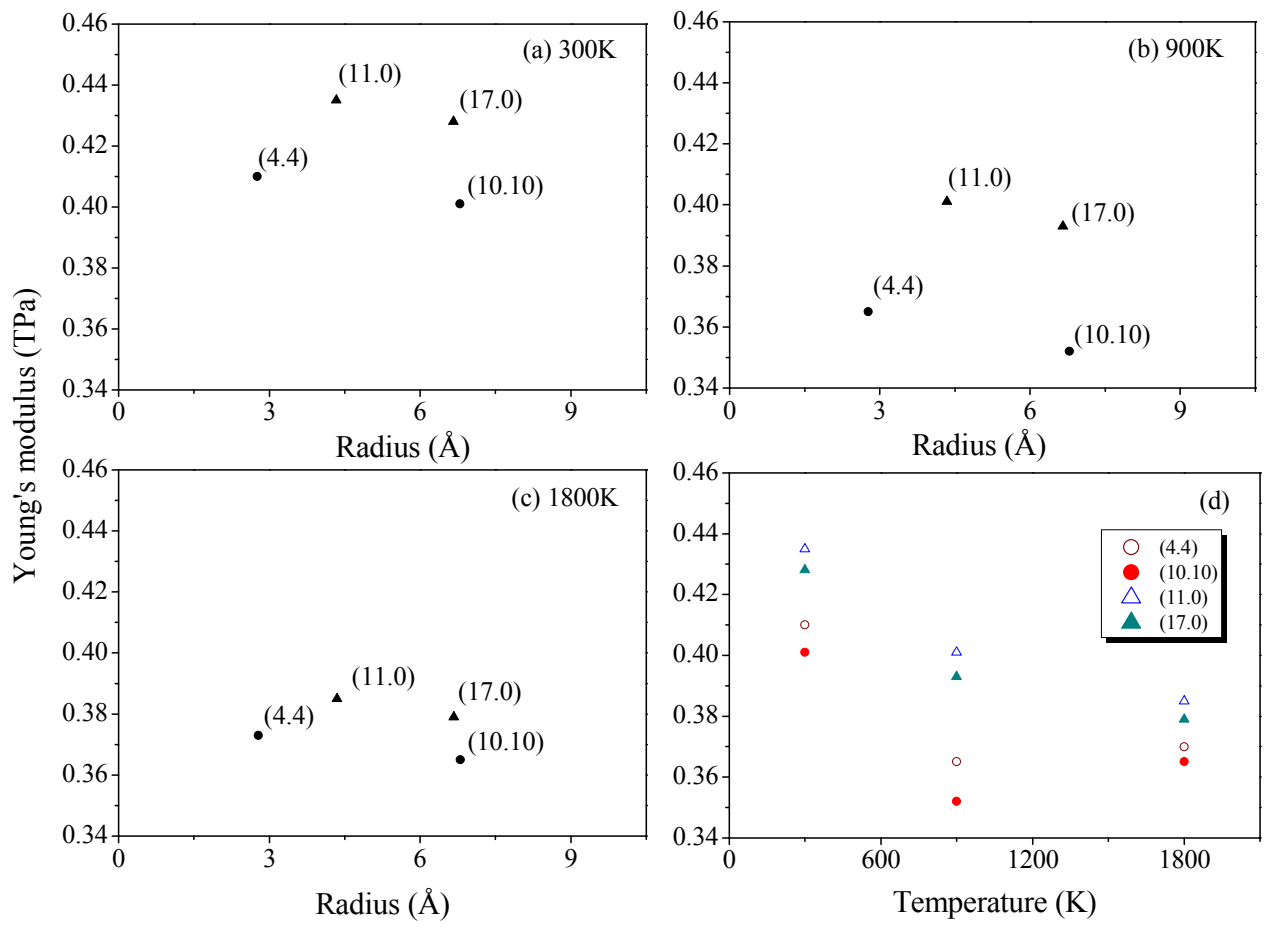

**Fig. 7.**